# Cross-Layer Extended Persistent Timeout Policy for SCTP and DSDV

M. Issoufou Tiado, R. Dhaou and A.-L. Beylot

*ENSEEIHT – IRIT, 2 rue Camichel BP 7122, 31071 Toulouse Cedex France +33 56158 8306*

<u>Issoufou_tiado@yahoo.fr</u>, <u>Riadh.Dhaou@enseeiht.fr</u>, <u>beylot@enseeiht.fr</u>

**Abstract**

*Cross layer techniques applied to various protocols stacks provide fair information sharing between OSI model layers. The performance gains have been demonstrated for many studied systems within protocols interactions. The example is illustrative of the reliable transport protocols that use retransmissions to achieve that reliability function. The performance gains of the persistent timeout policy for the management of the retransmission timeout have been produce in some recent works when applying that persistent timeout policy only to reliable transport protocol. The goal was to give an appropriate behavior in response to a bad state of the wireless channel that occurs and temporally blocks the transmission of data. The channel state is given by the 802.11 link layer through cross-layer mechanism. In this paper, the persistent policy is extended to the network layer and is applied to a stack that uses a reactive routing protocol, namely the Destination Sequenced Distance-Vector (DSDV) protocol that also generates additional periodic traffic regardless to the channel state. We are measuring the influence in terms of performance gains of the extended persistent policy because of the additional periodic signalization messages deriving from the used routing protocol.*

*After the introduction in section I; Section II of this paper presents an overview of the Stream Control Transmission Protocol (SCTP). Section III describes the behavior of the DSDV protocol. Section IV presents the extended persistent timeout policy principle and Section V presents the simulation results used to compare the using of the traditional and the extended persistent timeout policies applied to the same protocol stack using SCTP and DSDV.*

## 1. Introduction

Ad hoc networks consist of a set of wireless mobile nodes. The network is mainly characterized by the unpredictable nature of the channel state which can alternate from "good" to "bad" and vice versa asynchronously. The cross layer techniques allow a particular layer of protocols to take into account an information provided by another layer. The cross layer adaptation principle is used in [1] to design many mechanisms that allow an application to adjust its behavior by self configuring becoming necessary to meet the constraints that occurs in the network wireless channel. For example, the adaptation mentioned here allows the reliable transport protocol to differentiate different types of packet loss (losses due to congestion or due to channel errors) before invoking congestion control and flow regulation mechanism [2]. By this fact of adaptation using cross layer, many other implementations show the performance gain that can be achieve such as the persistent timeout management policy in [3]. Reliable transport protocols such as TCP [4][5][6] and SCTP [7][8][9] are subjected to some works that aim to improve their performances. In [3], the authors suggest that instead of using the non persistent retransmission policy in response to a bad state of the wireless channel that bring to the lack of acknowledgements, from the cross layer model in which the link layer provides the state of the wireless channel in the environment subsystem [10][11], the use of a persistent retransmission policy will generate performance gains when the bad state of the channel temporarily blocks the sending of data.

In the precedent work [3], the stack of protocol uses DSR protocol at the network layer, which is a reactive protocol and do not generate signalization messages. The basic idea of the extended persistent policy in this present work is to study the behavior of the system when using a proactive protocol such as DSDV which generates additional traffic without taken into account the state of the channel. Even with the fact that DSDV does not use retransmissions; the interest of the extension of the persistent policy is inherent to the protocol behavior that is presented later in the following section III.

After the presentation of SCTP and DSDV, the extended persistent policy is presented before the performance evaluation by simulation in ns-2 environment. A comparative study between the two retransmission policies of SCTP is conduct by using a proactive protocol at the routing level.

## 2. Overview of the traditional retransmission timeout policy of SCTP

The traditional Retransmission TimeOut Management Mechanism (RTO-MM) is used by SCTP in two cases: at the end of an association or for the management of errors that occur during the transmission.

Management of the end of an association at the sender level

When at the sender level the application sends the SHUTDOWN message to SCTP, it enters and remains into the SHUTDOWN-PENDING state until the transmission of the data waiting to be sent. Upon the reception of the acknowledgment of the last sent data, SCTP sends a SHUTDOWN to its opposite, starts the T2-shutdown timer and then passes into the SHUTDOWN-SENT state. When the sender node timer expires, another SHUTDOWN is sent, and so on until the maximal number of attempts is reached.
The reception of the SHUTDOWN-ACK by the sender triggers the T2-shutdown timer to be stopped, before the node forwards a SHUTDOWN-COMPLETE message, and then removes the current association.

Management of the end of an association at the receiver level

When the destination node receives a SHUTDOWN, it sends a SHUTDOWN-ACK, starts its T2-shutdown timer and goes into the SHUTDOWN-ACK-SENT state. Upon the expiration of the receiver's timer, another SHUTDOWN-ACK is sent, and so on until the maximum number of attempts. If the destination node receives the SHUTDOWN-COMPLETE message, it checks if it is in the SHUTDOWN-ACK-SENT state, than stops the T2-shutdown timer, sends a SHUTDOWN-COMPLETE, and deletes the association.

The proposed optimization

The sender and the receiver use the traditional policy to manage the two T2-shutdown timers. This policy allows the retransmission timer to take increasing values multiple of the previous to limit the number of attempts. By this fact, the extended persistent policy is applied to the management of the end of an association in response of a bad channel condition which results in the absence of a SHUTDOWN-ACK packet in response to a SHUTDOWN packet.

Transmission errors management

When errors occur during the transmission of a packet, SCTP uses the RTO-MM to ensure the reliability of the transmission. Upon the expiration of the retransmission timer or the reception of an SACK indicating that the data have not been received, the retransmission mechanism is triggered. To reduce the risk of the network congestion, the frequency of retransmissions is limited by the fact that the retransmission timer is adjusted based on the estimation of the "round trip delay" when the loss of messages increases. By the traditional RTO-MM, the timer takes a value multiple of the previous for each unsuccessful attempt. As SACK messages generate more retransmissions than the timer expiration in active association having a fair and consistent transmission of data, the rule of four SACK is used to reduce the possibly unnecessary retransmission. By this rule, the retransmission is triggered when the node receives four SACK indicating the loss of data and avoiding retransmission for cases of reordering. The use of the RTO-MM in this case of transmission errors management brings to its replacement by the actual extended persistent policy during "bad" state of the channel. For this proposed optimization, an evaluation of the contribution of the extended persistent policy must be performed.

## 3. Overview of DSDV

The classical Bellman-Ford routing algorithm is applied to the ad hoc networks with the design of the DSDV routing protocol [12]. The need of conveying the information through the network allows the mobile nodes to establish free collaboration for routing information. To achieve this objective, every mobile node maintains a routing table containing the list of all available destination nodes in the network. Each entry of the table contains a destination, the number of hops to reach the destination, the sequence number of the entry used to distinguish new routes and to avoid loops. The particularity of the DSDV protocol that is put under performance test in this paper is its behavior making a node to transmit periodically or if a significant change occurs in the network, its routing table to its immediate neighbors. DSDV does not use the retransmission mechanism and do not manage retransmission timer, but the periodic transmission of routing table of a node generate an additional traffic that can be take into account when the channel state is unfavorable. The extended persistent policy of the SCTP is proposed to avoid unnecessary traffic at the transport layer, and is extended to the network layer. Taking into account the DSDV activity to avoid unnecessary traffic is considered under the extended persistent policy when the channel state is bad, particularly because the routing tables are updated according to a preset time or according to a particular event that derived from the network change. Thus the DSDV node can use the "full-dump" update mechanism or the "incremental" update mechanism. When using the full-dump, a node sends the entire routing table to its neighbors, in contrary, with the incremental mechanism, only entries of the routing table whose sequence number has changed are sent. The incremental mechanism allows to avoid extra traffic, particularly when the network is relatively stable. But fast mobility of nodes can cause fast changing inside the network, and by this fact, full-dump will be more frequent. Also, the nodes use the settling time of routes by delaying the transmission of routing tables to eliminate updates that would occur if better routes are found very soon.

## 4. Extended Persistent Policy

The SCTP Persistent Timeout Policy
At the SCTP sender side, during the transmission of data or during the transmission of control messages, the expiration of the retransmission timer or the reception of a SACK indicating the lost of a packet trigger the retransmission of the packet. The RTO-MM retransmission policy used by the SCTP sender node is that the retransmission intervals grow up by taking for every next attempt, a value multiple of the last taken. This traditional RTO-MM of SCTP is similar to the operation of a non-persistent MAC level protocol that requires a non-persistent random timeout when the channel is busy before the next attempt. The SCTP persistent timeout policy proposed in [3] is derived from the adaptation of the behavior of persistent MAC level protocols that continue to observe the channel and transmit the frame when it is free. For the cross layer adaptation needs, the SCTP processes the channel state information provided by the 802.11 link layer by continuously observing that state and retransmit the pending packet at the next favourable change, when the channel status becomes favourable. The authors of [3] show the advantages of that persistent policy in terms of the latency gain, the energy consumption gain, the theoretical maximum throughput gain and the unsuccessful message emissions gain.

The extension of the Persistent Policy
When using the extended persistent policy at the network layer during an unfavorable state of the channel, the node must delayed the transmission of routing table and must take into account the duration of that bad state in the determination of the topology change. For example, if a running timer used to determine the interval between the sending of two consecutive update packets from a neighbor expires, the node must ignore this expiration until the next favorable change of the channel state. When by using cross layer mechanism through the environment subsystem the 802.11 link layer makes available the state of the channel, it becomes possible for both the SCTP at the transport layer and the DSDV at the network layer to process this information, by taking into account the occurrence of a bad state of the channel. The SCTP has to wait for the favorable change before sending its segments, the DSDV has to delay the transmission of data packet and to delay the periodic transmission of routing tables. The DSDV has also to suspend the update of the status of a neighbor that can be seen as a change that occurs in the network topology. Thus DSDV must not perform nor the full-dump neither the incremental route update mechanism during the bad state of the channel. The protocol must also continuously observe the channel state until the next favorable change, before recovering its normal behavior. The basic idea is that it will become possible to improve the performances of the global stack by avoiding unnecessary additional traffic which will bring to more energy consumption. It is interesting to evaluate by simulation if this extended persistent policy will generate more theoretical maximum throughput and avoid additional latency than the actual system using the RTO-MM at the transport layer with no processing of the channel state information. The difference between the system describe in [3] and the system that is being evaluated in this paper, comes from for the later work, the use of a reactive protocol at the routing level and for the actual work, the use of a proactive protocol. Because mainly, in the first case, the DSR protocol will start the transmission process upon receiving the packet from the SCTP, and if the SCTP is silent, the DSR will not generate additional traffic. In contrast, when using the DSDV protocol at the network layer for routing the packet provides by SCTP, even when the SCTP is silent, the DSDV will generate additional traffic because of its proactive nature. It is why it becomes interesting to measure the performance of that second system different from the first one. And as the immediate channel is questioned, the DSDV bring additional interest with the Distant Vector algorithm it implements, which result mainly to the exchange between neighbors, by using for all the routing traffic, the immediate channel.

Principle of continuing the evaluation of the channel state
As describe in [3], the extended persistent policy will bring the nodes to the same silent behavior during the bad state of the wireless channel. But the channel state is generated by the link layer particularly by the evaluation of the Signal to Noise Ratio (SNR), the Bit Error Rate (BER), the loss rate of packets sent or received, the retransmission rate, all these parameters are calculated from the activity of the physical layer. For the necessary need of continuing to evaluate the channel state, in [3], the following three events have been considered: (1) the use of the ambient activity if the network is in a high load state, (2) the use of pro-active routing protocol that uses periodic control messages to establish network connectivity or when a loss of the connectivity occurs (as in the case of poor channel), and (3) the use of a compensation mechanism when the environment subsystem triggers a request to update a road referenced in one of the pending packets to be sent. The environment subsystem will use the channel quality generation principle to provide the information about the wireless channel state. But for the current work for the used protocol stack and the extended persistent policy, that scheme must be modify to take into account the persistent behavior used for the proactive routing protocol. Thus, only the first and third case can be used to ensure the continuity of the evaluation of the channel state.

## 5. Performance Evaluation by Simulation

### 5.1. Description of scenarios

The objective of the simulation is to determine the influence of the extended persistent policy in contrast of the RTO-MM. Particularly, it is to evaluate the performance gain that can be achieve in the study of the behavior of the global system in the following two modes: (1) the traditional timeout policy of SCTP and the traditional behavior of DSDV and (2) the persistent timeout policy of SCTP extended to DSDV protocol. The simulated protocol stack is composed of SCTP at transport layer, the DSDV protocol and IP protocol [14] used at the network layer, the IEEE 802.11 [15] protocol for lower layers. The scenario is the same as in [3]. The simulation is run in ns-2 environment [16] with a node transmitting a Constant Bit Rate (CBR) traffic to a receiving node within a variable channel state. The inter-arrival time of messages during peak is set to 0.1 second. For each case, the policies are evaluated with segments of 1500 bytes. For the state model of the wireless channel, the Randomly Alternated Cut Model (RACM) is used to match different times of the evolution of the traffic, and allows evaluating the influence of the persistent policy. From 200th second of the simulation, an interval of availability randomly follows a period of unavailability of the channel. The durations of the availability and unavailability of the link are randomly switched alternately between 20 and 100 seconds. The interval of 20 seconds is chosen to exceed the first retransmission attempt of SCTP. The simulated scenarios are identical to help to make the comparison. Each scenario is evaluated 20 times with the influence of the random break, and the results are summarized by grouping the durations of breaks, and by calculating the average of each evaluation for each given interval value.

### 5.2. Basis of the interpretation of the results

In [3], the comparison criterions of the simulation are the latency, the theoretical maximum throughput, the unsuccessful message emissions and the energy consumption. The same criterions are considered in this case to evaluate the performance gain of the extended persistent policy compared to the traditional RTO-MM. The latency can be defined as the delayed time before sending the data when the channel becomes available. The theoretical maximum throughput comes from the intuition by which if the new policy reduces the latency, the node can inject additional traffic. The energy consumption criterion derives from the law of Gallager [17]: the energy of a mobile node can be converted into the number of bits that the node can send before exhausting totally that energy. Thus, the retransmission attempts of packets falls under the energy law of Gallager. In the curves below, these three criterions take an average value of the results obtained during all the run scenarios.

### 5.3. The influence of DSDV on the latency curves

The average latency criterion is used to compare the RTO-MM and the extended persistent policy applied on the stack using SCTP and DSDV a proactive routing protocol. The curves in Figure V.1 below as in the previous study that put into association the reactive routing protocol DSR at the network layer to the SCTP at the transport layer, confirm the improvement provide by the extended persistent policy in terms of average latency calculates and groups according to the break duration of randomly alternated cut model. The results shown in these curves can be explain by the fact that when the channel becomes available after a period of break, the stack of protocol take different time before sending data according to the used policy. The traditional RTO-MM combined with the normal behavior of DSDV reflects the behavior of SCTP and DSDV, namely, the waiting of the expiry of the timer and the sending of the heartbeat packets by the SCTP when the channel becomes available before sending its data. This behavior is combined with the route establishment of the DSDV protocol, because when the bad state of the channel take a period of time, the routing structure of DSDV are modified due to the lack of periodic message coming from neighbor. DSDV protocol takes a non negligible amount of time to set route before sending its data.

When using the persistent policy, SCTP and DSDV suspend their periodic activity and wait for the next favorable change of the channel state before sending the pending data. The two protocols mechanisms are reactivated when the channel becomes available. And Thus depending to the modifications that occur in the DSDV routing structure and depending of the success get by the SCTP to ensure the availability of the destination by exchanging Heartbeat messages regardless to the timer that steel be suspend, the latency is improved by the extended persistent policy as shown on the curves. DSDV use limited route discovery without taken into account the entire topology of the network and limiting its exchange towards its immediate neighbors. In conclusion for this criterion of latency calculation, by intuition and according to the results provide by the curves, the extended persistent policy shows less latency than the traditional RTO-MM.

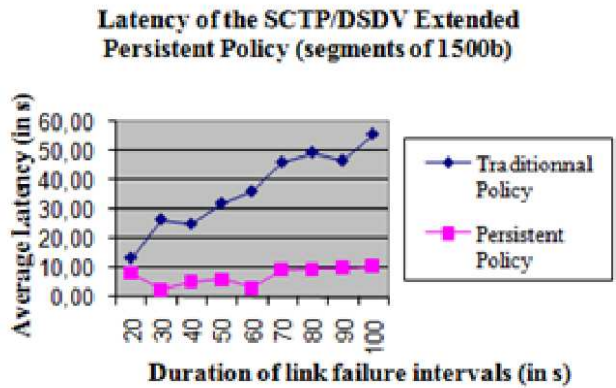
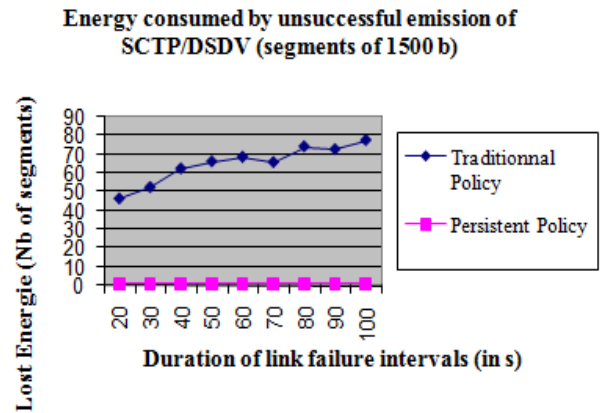

**Figure V.1. Average latency of SCTP and DSDV.**

**Figure V.2. Unsuccessful emissions of SCTP and DSDV.**

### 5.4. The influence of DSDV on the Energy consumed by unsuccessful emissions

The SCTP/DSDV stack has a particular interest, because of the periodic emission that characterized the two protocols in addition of the traditional response given by SCTP in response to the lack of acknowledgements (the RTO-MM). When a bad state of the channel occurs, this behavior does not change in the traditional RTO-MM and generates more unsuccessful emissions and thus more energy consumption that is not negligible as a mobile node has a finite energy. The persistent policy extended to the DSDV protocol in its exchange between its neighbors while running the distant vector algorithm brings the protocol to suspend its periodic emission and observe the next favorable change of the channel before performing the proactive route discovery. The SCTP is subjected to the same behavior. It suspends the periodic activity and observes the next good change that occurs in the channel state before restarting its Heartbeat message exchange and its data transmission. Also during the persistent time, the retransmission timer is managed differently to the traditional way, by steel being suspended in opposition of its management in the RTO-MM. All the suspended activities of the two protocols are restored upon the occurrence of the good state of the wireless channel. This new behavior brings by the extended persistent policy over the all stack provides less energy consumption than in the traditional RTO-MM as shown by the curves of the figure V.2 below. Even with the fact that in the traditional policy the DSDV generates less overhead by not trying to organize the entire network, but limiting its exchange only between its neighbors, the extended persistent policy improves the energy consumed by unsuccessful emissions criterion.

### 5.5. The influence of DSDV on the Percentage of unsuccessful emissions compared to the data transferred

The RTO-MM policy and the extended persistent policy can be compared on the base of the amount of transferred data relatively to the unsuccessful emission when the transmission of data is subjected to the vagaries of the channel. The vagaries of the channel are built in the randomly alternated cut model. The ratio that gives the percentage of unsuccessful emissions compared to the amount of the transferred data highlights the contribution of the persistent policy. The difference between the curves in the figure V.3 below comes from the traditional RTO-MM that generates more unsuccessful emissions than the extended persistent policy, and as the extended persistent policy allows to reduce the latency, the stack of protocol injects more data. Bringing the injected data in percentage of unsuccessful emissions, the extended persistent policy takes more advantage over the traditional RTO-MM. For example, for the interval of 100 seconds of unavailability of the channel, the extended persistent policy provides a better ratio than the traditional RTO-MM, with 0.8% for the first policy versus 1.4% for the second. The difference comes from the periodic emission of the Heartbeat messages and the periodic activity of DSDV towards its immediate neighbors. The long duration of the bad state of the channel impacts the DSDV structure which needs to re-establish route when the channel becomes available. This can brings more latency and thus reduce the amount of the transferred data, as the time of the simulation is delimited.

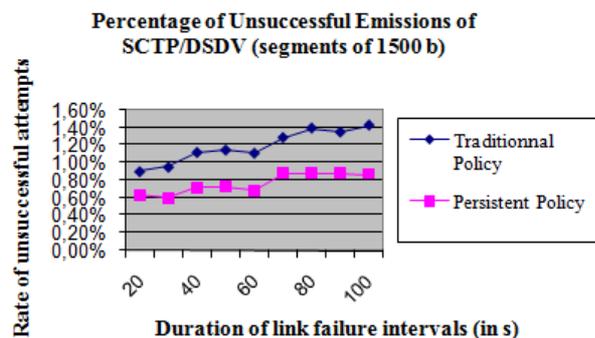

**Figure V.3. Percentage of the unsuccessful emissions by transferred data volume of SCTP and DSDV.**

## 6. Conclusion

The extended persistent policy is a cross layer mechanism applied to the SCTP/DSDV protocol stack in this paper. It is built as a better response to the occurrence of the vagaries of the wireless channel, particularly when the bad state temporally blocks the transmission of data, making the SCTP protocol invoking the traditional RTO-MM that generates unsuccessful emissions in addition with the unsuccessful emission of the periodic Heartbeat messages. Equally, when the bad state of the wireless channel occurs, it makes the DSDV generating unsuccessful emissions of periodic messages towards its neighbors used to establish route within the distant vector algorithm. By this way, the extended persistent policy reduces the sending latency, improves the amount of data transferred and generates less energy consumption during bad states of the channel.

The extended persistent policy is based on the channel state provides by the environment subsystem. An additional work to conduct is to measure the behavior of the entire stack under the stated policy of continuing the assessment of the channel state. It is also interesting to measure the improvement of the extended policy on other routing protocols, to study interactions between the protocols of the studied stack and to make the environment subsystem richer.